\documentclass[preprint,3p,11pt,onecolumn]{elsarticle}
\usepackage{natbib} % Required for Elsevier-style bibliography
\usepackage{amssymb}
\usepackage{amsmath} % Include the amsmath package
\usepackage{caption}
\usepackage{graphicx}
\usepackage[colorlinks,allcolors=blue]{hyperref}
\usepackage{color}

\journal{Additive Manufacturing}

\newcommand*\diff{\mathop{}\!\mathrm{d}}
\usepackage{siunitx}

\begin{document}

\begin{frontmatter}



\title{Structural fluctuations in thin cohesive particle layers in powder-based additive manufacturing}

\author{Sudeshna Roy\footnote{Equal contribution.}$^{1*}$}
\author{Hongyi Xiao\footnotemark[1]$^{1,2}$}
\author{Vasileios Angelidakis\footnotemark[1]$^{1}$}
\author{Thorsten P\"oschel$^{1}$}
\address{
1. Institute for Multiscale Simulation, Friedrich-Alexander-Universit\"at Erlangen-N\"urnberg, Erlangen, Germany.
\\ 2. Department of Mechanical Engineering, University of Michigan, Ann Arbor, USA}
% \affiliation{organization={Institute for Multiscale Simulation, \\ 
% Friedrich-Alexander-University Erlangen-Nuremberg, \\ 
% Cauerstrasse 3, 91058 Erlangen, Germany.}}
\cortext[correspondingauthor]{Correspondence: sudeshna.roy@fau.de (S.R.)}
% \cortext[equal]{Equal contribution}

\begin{abstract}
Producing dense and homogeneous powder layers with smooth free surface is challenging in additive manufacturing, as interparticle cohesion can strongly affect the powder packing structure and therefore influence the quality of the end product. We use the Discrete Element Method to simulate the spreading process of spherical powders and examine how cohesion influences the characteristics of the packing structure with a focus on the fluctuation of the local morphology. As cohesion increases, the overall packing density decreases, and the free surface roughness increases, which is calculated from digitized surface height distributions. Local structural fluctuations for both quantities are examined
through the local packing anisotropy on  the particle scale, obtained from Vorono\"{\i} tessellation.
The distributions of these particle-level metrics quantify the increasingly heterogeneous packing structure with clustering and changing surface morphology.

%The spatial inhomogeneity of the packing density is calculated within each powder layer, in addition to the local structural anisotropy, which is quantified using a method based on Vorono\"{\i} tessellation. The surface roughness of each layer is characterized to identify the influence of cohesion on the surface texture. Based on the statistics of the density, local anisotropy, and surface roughness, we quantify the increase of structural heterogeneity as cohesion increases. In particular, the distribution of local anisotropy widens, and the layer surface roughness increases with changing morphology.
\end{abstract}




\end{frontmatter}


\section{Introduction}
% Powder-based additive manufacturing, such as powder bed fusion, has attracted significant attention \cite{bhavar2017review,vock2019powders,chen2022review} due to its application in fast prototyping and for the production of parts with high geometrical flexibility. This branch \TP{which branch?} of methods allows fast near-net-shape \TP{DEFINE} production with minimal support structure \TP{powder method do not need any support structure}, saving time \TP{why? evidence?} and reducing material waste \TP{produces much more waste than methods that need support structure}. 
Powder-based additive manufacturing techniques, like powder bed fusion, have garnered considerable interest \cite{bhavar2017review,vock2019powders,chen2022review} for their ability to facilitate rapid prototyping and the production of highly customizable parts. These methods enable efficient manufacturing by minimizing the need for material removal and extensive support structures, which in turn reduces production  time and material waste.
% However, a more extensive adoption of this technology is still lacking \TP{90\% of all applications in industry are powder based} due to limitations in process quality control. 
However, the quality and efficiency of powder-based techniques are far from ideal. Non-uniform powder packing during spreading is one of the major issues that limit the range of available powder materials and impair printing quality. Various types of structural defects in the deposited powder layer have been observed, which strongly correlate to defects in sintered parts \cite{cunningham2017analyzing,gong2015influence,mostafaei2022defects}. 
Since commonly used particle sizes are far below 100\,$\mu$m, cohesion between particles can impair the spreading and deteriorate the quality of the powder layer through reduced powder flowability and cohesion-induced powder clustering. Understanding the influence of cohesion on spreading requires detailed measurement of the packing structure under various levels of cohesion, which is expensive and difficult to obtain experimentally \cite{gordon2020defect,parteli2014attractive,fang2022process,schmidt2020packings}. Characterizing a thin particle layer is also challenging, as most of the existing metrics are meant for bulk characterization.

%In additive manufacturing, the efficient and precise spreading of powdered material is a crucial process that significantly impacts the overall quality and integrity of the final product. 
One prominent method utilized to understand and design the powder spreading process is the Discrete Element Method (DEM), which is a particle-based simulation technique that computes particle trajectories from the interaction forces. Various DEM-based studies have investigated powder spreading with the aim of improving the quality of the powder layer \cite{Parteli:2014,nasato2020influence,shaheen2021influence,he2021combined,fouda2020study,haeri2017discrete,chen2019powder,meier2019critical}. Simulations can reflect behaviors of real powders~\cite{meier2019modeling,he2020linking} during spreading as they can be calibrated by experiments of powder flowability \cite{zhang2020discrete,shaheen2021influence,han2019discrete,meier2019modeling}, which allows detailed studies of the influence of process parameters and powder properties on spreading. For example, Parteli and P\"oschel \cite{parteli2016particle} showed in simulations that a fast spreading process increases the surface roughness of a cohesive powder layer for roller spreading. Nasato et al. \cite{nasato2021influence} found that small frequency and amplitude of a vibrating recoater lead to low powder bed porosity. Non-spherical powders with realistic particle shapes were also considered when investigating how the recoating velocity influences the bed porosity~\cite{parteli2017particle,nasato2020influence}. Shaheen et al. \cite{shaheen2021influence} showed that powder layer defects are more likely to occur with higher particle rolling and sliding friction. 

In these studies, the prerequisite of establishing the relation between process and material parameters and the layer quality is a detailed and informative characterization of the packing structure, which can be challenging for cohesive particles due to effects like clustering. While the global packing density is informative and widely used, it does not contain information of how the particles are spatially arranged. Therefore, the spatial fluctuation of the packing structure is also important, especially for highly cohesive powders where the packing tends to be heterogeneous \cite{chen2022review}. To this end, local density is often calculated using binning and coarse-graining where the averaging length must be chosen \cite{meier2019critical,shaheen2021influence}. Metrics based on the Vorono\"{\i} cell volume can also be used, which does not require hand-picking an averaging length scale. For example, Phua et al. used Vorono\"{\i}-tessellation of particles in 3D to calculate the average packing fraction of powder layers\cite{phua2023understanding}. However, examining the global distribution of the Vorono\"{\i} still does not offer the complete picture of how density fluctuates. Here, we adopt a Vorono\"{\i}-tessellation based method \cite{rieser2016divergence} to quantify local structural anisotropy, which is an inherent property of non-crystalline packing of particles and is associated with critical mechanical properties in disordered packings, such as jamming \cite{rieser2016divergence}, plasticity \cite{richard2020predicting}, and shear band formation \cite{xiao2020strain,harrington2018anisotropic,harrington2020stagnant}. This method does not require choosing a density threshold to identify voids and it yields a meaningful distribution of local anisotropy in a deposited powder layer, based on which the hetereogeneity of the packing can be quantified.

 The surface roughness of the deposited powder also plays a crucial role in determining the functionality and aesthetics of the final product. Achieving the desired surface finish is essential for optimizing performance and ensuring consistent product quality. Surface roughness, similar to density, is also influenced by the interplay between process parameters \cite{parteli2017particle,haeri2017discrete,mussatto2021influences,chen2020packing} and material properties such as cohesion, particle size distribution \cite{he2021combined} and particle shape \cite{nasato2020influence, nan2018jamming}. In particular, cohesion strongly influences the surface roughness during spreading. The powder bed surface roughness increases with cohesion due to powder agglomeration and particle removal caused by particle-to-blade cohesion during spreading \cite{meier2019critical,he2020linking}. In DEM simulations, the surface roughness is typically evaluated by measuring the local surface height determined by the maximum vertical coordinate of the powder bed and monitoring its spatial variation. Using this variation as a metric of uniformity, surface roughness is calculated as the mean deviation of surface height from the powder bed average height \cite{he2021combined, meier2019critical}. Experimentally, the surface height can be determined using optical 3D digital microscopy \cite{mussatto2021influences} or high-speed laser profilometry \cite{chen2020packing}. Surface roughness can be quantified either using planar profile measurements in two dimensions\cite{nasato2020influence,parteli2017particle} or areal measurements in three dimensions \cite{meier2019critical,angelidakis2021shape}.
 
While metrics like the standard deviation of the global surface height distribution is informative, it does not offer a complete description of the surface profile. In this study, we evaluate the skewness and the kurtosis of the height distribution calculated using an efficient digitization method \cite{angelidakis2021shape}. These characteristics offer further insight into the presence of local outliers in surface roughness and the extent to which they deviate from the mean surface plane. We also address the problem that for a given set of surface height values, the distribution cannot well describe the local fluctuations because a spatial rearrangement of the height values does not change the distribution. This is similar to the aforementioned problem that the global packing density cannot sufficiently describe the heterogeneity of the packing. To this end, we quantify the spatial fluctuations of the free surface height of the powder layer through a coarse-graining approach to calculate the squared local spatial gradient of the Vorono\"{\i} cell-averaged height. This quantity again yields a meaningful distribution that can be described by a single parameter, quantifying the height fluctuations.
\section{Model}
\subsection{Numerical Setup}
\begin{figure}[!htb]
    \begin{center}
        \includegraphics[width=0.65\columnwidth,trim={0cm 0cm 0cm 2.0cm},clip]{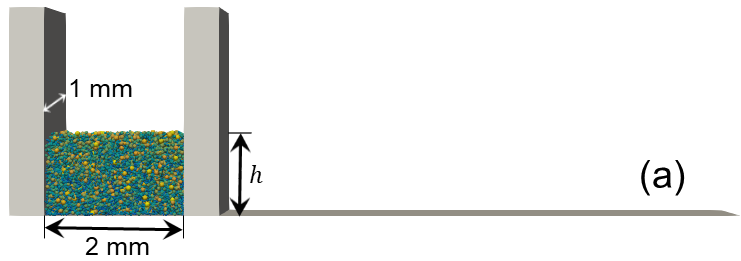}
        \hspace{4pt}
        
        \vspace{0.6cm}  
        \includegraphics[width=0.65\columnwidth,trim={0cm 0cm 0cm 2.0cm},clip]{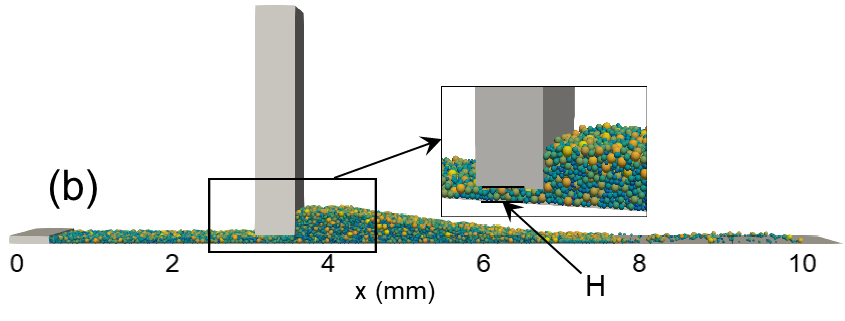}
        \hspace{4pt}
        
        \vspace{0.6cm}
        
        \includegraphics[width=0.65\columnwidth,trim={0cm 0cm 0cm 2.0cm},clip]{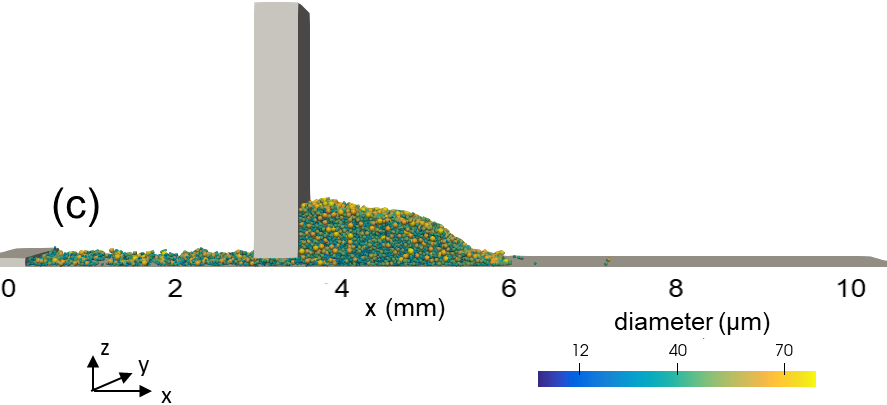}
    \end{center}
    \caption{Numerical setup for powder spreading on a planar substrate during spreading showing (a) initial configuration and spreading setup for (b) cohesionless and (c) cohesive powders. %\TP{(Bond number not yet defined. I suggest to define the Bond number in Sec. 2.3 with a numbered equation and cite it here.)} 
    %with sliding friction coefficients $\mu=0.10$.
    }
    \label{fig:schematic}
\end{figure}

We employ DEM to obtain particle-scale information on powder layers created by a spreading process, using \texttt{MercuryDPM} \cite{weinhart2020fast}.  The simulation setup is shown in \autoref{fig:schematic}. The powder is spread by a blade tool, moving at constant velocity $v_T$ along the spreading direction, $x$ \cite{roy2023effect,roy2022local}. We simulate a small slice of the powder bed of length $10$ mm in the $x$-direction and width $1$ mm in the lateral $y$ direction where periodic boundary conditions are applied. For the subsequent analysis, we consider the range $0\le x \le 7\,\text{mm}]$.
It is assumed that the substrate is flat and the coefficient of friction between the wall and the particles is equal to that of the particle-particle interaction. A log-normal particle size distribution is considered with mean particle diameter $D_{50} = 37 \, \mu$m, $D_{10} = 24 \,\mu$m, and $D_{90} = 56 \,\mu$m. 
% \TP{(This is a contradiction. A log-normal distribution cannot be characterized by smallest and largest particles)} 
The particles are initially generated in front of the spreader tool at $(x,y,z) \in [0.5,2.5]$ mm $\times [0,1]$ mm $\times [0,h]$ as shown in \autoref{fig:schematic}(a), filling a total bulk particle volume of $0.75$ mm$^3$ which is sufficient to create a powder layer of $10$ mm length, $1$ mm width, and tool gap $H=100 \, \mu$m, where the tool gap is defined as the gap between the base of the blade and the substrate, as shown in \autoref{fig:schematic}(b).
% \TP{(``tool gap'' is not defined)} 
% \TP{(Why write so complicated? - Ask someone to draw a sketch due to your description. Can this really be understood?)} 
%After the particles settle under gravity and the system relaxed, 
The spreading process starts at a constant velocity $v_T=10$ mm/s with initially all particles at rest. It ends when the blade arrives at the end after $1.2$ s,
% \TP{(this is surprising: x-length is 2 mm, velocity is 10 mm/s and time is 1.2 s???)}, 
% \SR{x-length is 10 mm, velocity is 10 mm/s, therefore total spreading time 1 s and 0.2 s additional time for depositing the particles}
and the simulation ends at time $1.5$ s once the system is relaxed again, i.e., when the kinetic energy is sufficiently low. 
% \TP{(one more length. - put all of these lengths to Fig. 1 such that the setup can be understood.)}

 \subsection{Contact models}\label{app:non-linear contact model}

\subsubsection{Hertz-Mindlin visco-elastic contact model}
The visco-elastic Hertz-Mindlin contact model (no-slip solution) \cite{thornton2013investigation,thornton2015granular}
% \TP{(not true, see Eq. (4))}
is employed to calculate the normal and tangential elastic contact forces between particles, respectively. The normal force for the Hertz visco-elastic model is given as
\begin{equation}
    \vec{F_n} = \min\left(0, -\rho \xi^{3/2} - \frac{3}{2}A_n\rho\sqrt{\xi}\dot{\xi}\right) \vec{e_n}
    %F_n = 4/3 \cdot E^{*} \cdot \sqrt{R} \cdot \delta^{3/2} + A
\end{equation}
where $\xi = R_i + R_j - \vert\vec{r}_i-\vec{r}_j\vert$ is the compression of two interacting particles $i$, $j$ of radii $R_i$ and $R_j$ at positions $\vec{r}_i$ and $\vec{r}_j$ and $\vec{e}_n = (\vec{r}_i-\vec{r}_j)/\vert\vec{r}_i-\vec{r}_j\vert$ is the normal unit vector, $A_n = 5\times 10^{-6}$ s is the normal dissipative parameter, calculated as in \cite{muller2011collision}, considering a coefficient of restitution of $0.4$ for the characteristic blade velocity $10$ mm/s and
\begin{equation}
    \rho = \frac{4}{3} \, E^* \, \sqrt{R^*} 
    \label{eq:Hertz_rho}
\end{equation}
with the effective radius $R^*$.
The effective elastic modulus,
\begin{equation}
     E^* = {\left(\frac{1-\nu_i^2}{E_i} + \frac{1-\nu_j^2}{E_j}\right)}^{-1}
\end{equation}
depends on the elastic moduli and the Poisson ratio of the material of particles $i$ and $j$. 

We model the tangential viscoelastic forces 
% \TP{(what is a ``path-dependent calculation''? Is there a path-independent calculation? Simply write what the algorithm does.}
following the no-slip solution of Mindlin \cite{mindlin1949compliance} for the elastic part and Parteli and P\"oschel \cite{parteli2016particle} for the tangential dissipative constant $A_t \approx 2 A_n E^*$, which are capped by the static friction force between two particles.
The tangential force is given by
% \begin{equation}
%     \vec{F_t} = -\min \left[ \mu|\vec{F_n}|,  \int_{path}^{} 8 G^{*}\sqrt{R^* \xi} \,ds 
%     + \eta\sqrt{m^*G^*\sqrt{R^*\xi}} \, v_t \right] \vec{e_t}
% \end{equation}

% \noindent
% The tangential force is given by
\begin{equation}
    \vec{F_t} = -\min \left[ \mu|\vec{F_n}|,  \int_{path}^{} 8 G^{*}\sqrt{R^* \xi} \,ds 
    + A_t  \sqrt{R^* \xi} v_t \right] \vec{e_t}\,,
\end{equation}
with the friction coefficient, $\mu$, the effective shear modulus
\begin{equation}
    G^*=\left(\frac{2-\nu_i}{G_i} + \frac{2-\nu_j}{G_j}\right)^{-1}
\end{equation} 
which for particles of identical material simplifies to $G^*=\frac{4G}{2-\nu}$, and the tangential relative displacement of the particles, $ds$.

\subsubsection{Non-linear cohesive model}
To simulate particle cohesion, we incorporated adhesive forces described by the Johnson-Kendall-Roberts model \cite{johnson1971surface} (JKR) and attractive forces using a model for non-bonded van der Waals interactions \cite{parteli2014attractive}. The JKR adhesive force is computed as
\begin{equation}
    \vec{F}_\mathrm{JKR} = 4\sqrt{\pi a^3\gamma E^*} \; \vec{e_n}
\end{equation}
where $\gamma$ is the surface energy density and $a$ is the contact radius related to deformation, calculated using
\begin{equation}
    \xi = \frac{a^2}{R^*} - \sqrt{\frac{4\pi a\gamma}{E^*}}
\end{equation}
The maximum interaction distance at which the contact breaks under tension is given by \begin{equation}
    \xi_t = \frac{1}{2}\frac{1}{6^{1/3}}\frac{a^2}{R^*}
\end{equation}
The non-bonded van der Waals attractive force \cite{parteli2014attractive,hamaker1937london, eggersdorfer2010fragmentation} reads  
\begin{align}
    \vec{F}_\mathrm{vdW} = 
    \begin{cases}
        \frac{A_HR^*}{6D^2_\mathrm{min}} \, \vec{e}_n,& \text{if } \xi > 0\\ 
        \frac{A_HR^*}{6{(\xi-D_\mathrm{min})}^2} \,\vec{e}_n,& \text{if } -D_\mathrm{max}  \leq \xi \leq 0\\
        0,              & \xi < -D_\mathrm{max}
    \end{cases}
\end{align}
where $D_\mathrm{min} = 1.65$ \AA ~ is a parameter introduced to avoid a singularity \cite{parteli2014attractive}, $D_\mathrm{max}$ is the maximum interaction distance of the van der Waals interaction, which is set as $1~\mu$m \cite{parteli2014attractive} and $A_H$ is the Hamaker constant which relates to the surface energy density via 
\begin{equation}
    A_H = 24\pi D^2_\mathrm{min}\gamma\,.
\end{equation}
\subsection{Material Parameters}
% We perform this numerical study using a realistic \TP{(This is certainly not true. Some of the values in Tab 1 are not even reasonably defined. What does a variable with units kg/sec$^2$ describe? same foe kg/sec??)} particle model of Ti-6Al-4V. 
The powder spreading process is simulated considering a metallic Ti-6Al-4V powder. The material and simulation parameters can be found in \autoref{tab:material_parameters}. According to the experimental measurement of the angle of repose of $41^\circ$ and matched with the simulation results of Meier et al. \cite{meier2019modeling}, the surface energy of Ti-6Al-4V is $0.1$ mJ/m$^2$.
% \TP{(Why not add these parameters to the table where all other are?)} 
%\SR{Added this to the table}
To study the effect of particle cohesion on the powder quality, we simulate the powder spreading process for varying surface energy $\gamma$ from $0$ to $0.5$ mJ/m$^2$ in steps of 
 % \TP{(?? Do you mean ``in steps of''??)} 
 $0.05$ mJ/m$^2$. In general, for cohesive bonds, the surface energy $\gamma$ quantifies the energy associated with disrupting a bond between cohesive particles to create surface. Thus, varying $\gamma$ 
 %and not $k^\mathrm{coh}$ 
 is a meaningful representation of varying cohesion intensity between neighboring particles. We introduce the Bond number 
  \begin{equation}\label{Eqn:Bo}
 Bo=\frac{36\gamma}{\rho D_{50}^2g}
 \end{equation}
to characterize the ratio between interparticle cohesion and gravity,
 % \TP{(what is ``particle gravity''?)}
% the Bond number is employed, which is calculated as
where $D_{50} = 37\,\mu$m. 
 % \TP{($D^2_{50}$ is not defined)} 
 {Note that $Bo=54.5$ corresponds to the surface energy of the Ti-6Al-4V powder with $\gamma = 0.1$ mJ/m$^2$ \cite{meier2019modeling}. For the given particle and material parameters, the values of $\gamma\in\{0, 0.05, 0.1, 0.15, 0.2, 0.25, 0.3, 0.35, 0.4, 0.45, 0.5\}$ mJ/m$^2$} correspond to $Bo \in \{0, 27.2, 54.5, 81.7,$ $108.9, 136.2, 163.4, 190.6, 217.9, 245.1, 272.3\}$. 
% \TP{(Why do you need the rounded values?)}

\begin{table}[htb!]
\captionsetup{justification=centering}
\caption{DEM simulation parameters}
\label{tab:material_parameters}
\begin{center}
\begin{tabular}{l@{\quad}l@{\quad}ll}
\hline
\multicolumn{1}{l}{\rule{0pt}{18pt}%
variable}&\multicolumn{1}{l}{unit}&{value}\\[2pt]
                   \hline\rule{0pt}{12pt}\noindent
                   particle density ($\rho$)  & kg/m$^3$& $4430$\\
                   elastic modulus ($E$)  & MPa & $2.30$\\
                   % Dissipative constant ($A_n$)  & s& $5\times 10^{-6}$\\
                   Poisson's ratio ($\nu$)  &  -& $0.40$\\
                   sliding friction coeff. ($\mu$)  & -& $0.10$\\
                   particle diameter ($d_p$)  &$\mu$m& $12-79$\\[2pt]                   
\hline
\multicolumn{3}{l}{\rule{0pt}{18pt}\noindent 
                   additional parameters describing cohesion}\\[2pt]
                   \hline\rule{0pt}{12pt}\noindent
                   surface energy ($\gamma$)  & mJ/m$^2$& $0-0.5$\\ 
                   surface energy of Ti-6Al-4V & mJ/m$^2$& $0.1$\\ 
                   interaction distance ($D_\mathrm{max}$)  &$\mu$m& $1$ \\[2pt]
                   %{\color{red}Hamaker constant ($A_H$)}  & J& $0.2 \times 10^{-21}$\\                  
                              
\hline                   
\end{tabular}
\end{center}
\end{table}

\section{Local density characterization of the powder layer}
The quality of the produced powder layer is closely related to the packing density of the particles prior to sintering \cite{he2020linking}. The density of the layer can be quantified by the ratio of the volume occupied by particles and the total volume, $\phi=V_{\text{solid}}/V_{\text{total}}$. For sufficiently small $V_\text{total}$, $\phi$ can be considered as a local variable. Low packing fraction values indicate loose structures that are prone to defects in the final product. In general, a high packing fraction is desirable for high product quality.

% \TP{--------------------------------------------}

% \TP{I have to stop here for the moment with careful reading. I made many small but unimportant changes in the entire manuscript. I think we should talk before proceeding.}

The packing density is calculated locally for subsections of each layer to provide information about the spatial variability of voids throughout the layer. To this end, the local packing fraction is calculated for horizontal strips across the spreading distance $x$, of fixed width equal to $1$ mm.
The strip size is chosen to be sufficiently large so that it contains a representative number of particles and voids for the calculation of the packing fraction. The density calculations are performed using \texttt{YADE}, where a dedicated algorithm exists for density calculations \cite{yade:doc3}. 
%It should be noted that 
Alternative, high-resolution techniques have been proposed to calculate the density of granular packings based on the exact partial intersection volume between spheres and mesh elements \cite{strobl2016exact}.
\begin{figure}[htb!]
%       \begin{minipage}{0.5\textwidth}
        \centering
%        \fbox
        {\includegraphics[width=0.7\textwidth,trim={3cm 9cm 3cm 9cm},clip]{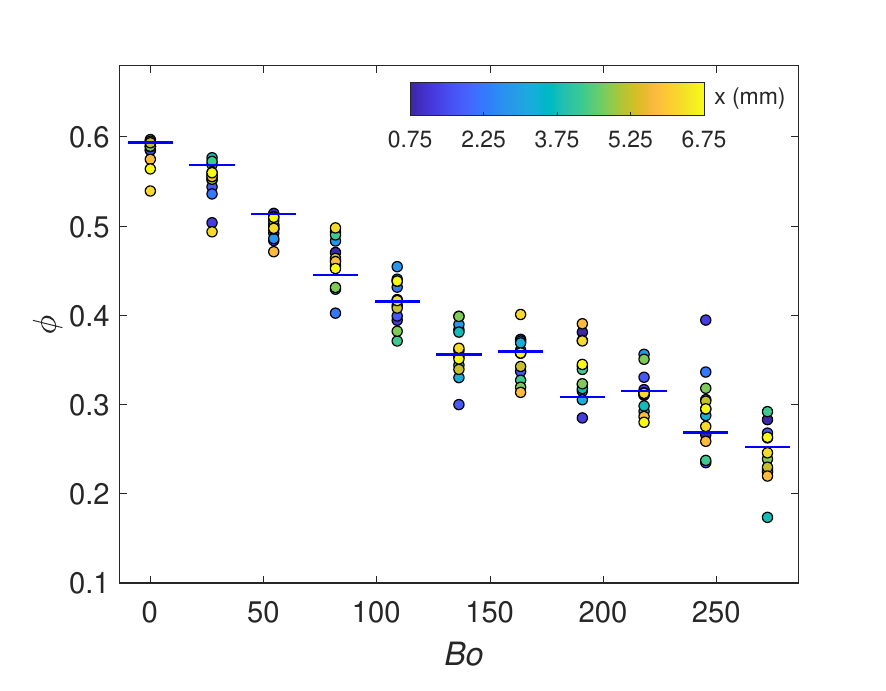}}
%        {\includegraphics[width=0.7\textwidth]{figures/packingFraction.png}}        
% \end{minipage}\hfill
\caption{Local packing fraction $\phi$ as a function of Bond number $Bo$. The horizontal lines note the global packing fraction value for each layer. The sample points are colored according to their distance $x$ from the starting point of spreading.}
\label{fig:packingFraction}
\end{figure}

Powder spreading leads to unstructured, inhomogeneous layers of material with spatially varying packing characteristics. This is the motivation behind calculating the packing density locally, for subsections of the layer along the spreading direction, aiming to explore the degree of density inhomogeneity within each layer. \autoref{fig:packingFraction} shows values of the packing fraction for various values of cohesion as a function of $Bo$. Evidently, cohesive materials lead to loose packings of the powder material, which is in agreement with previous observations \cite{he2020linking}. For increasing cohesion from $Bo=0$ to $Bo=272.3$, the mean packing fraction reduces by nearly 60\%, from $\phi\approx0.60$ to $\phi\approx0.25$, while the scattering of the values also increases slightly.
%for higher bond numbers. 
%The latter indicates that layers made of cohesionless powders $\left(Bo=0\right)$ demonstrate more homogeneous density spatially, while layers made of cohesive powders, e.g. $\left(Bo\approx30\right)$ exhibit more extensive scattering of $\phi$ (from $\phi\approx0.13-0.29$ in \autoref{fig:packingFraction}).
The reduction of the average packing fraction is gradual for increasing Bond number (within the studied range of values), i.e., no sudden transitions are observed between layers made of powders with similar Bond numbers. The data points in \autoref{fig:packingFraction} are colored according to their distance, $x$, from the starting point of spreading, where a clear trend is not observed, indicating that the degree of scattering does not correlate with the spreading distance, $x$. %While the binning size needs to be specified for this strip-based calculation, we better quantify the fluctuation with a local measure in the next section.
\section{Local structural anisotropy characterization of cohesive particle layers}
\label{sec:local_structural_anisotropy}
\subsection{Heterogeneous packing of cohesive particles}
\begin{figure}[!htb]
  \begin{center}
    \includegraphics[width=\columnwidth]{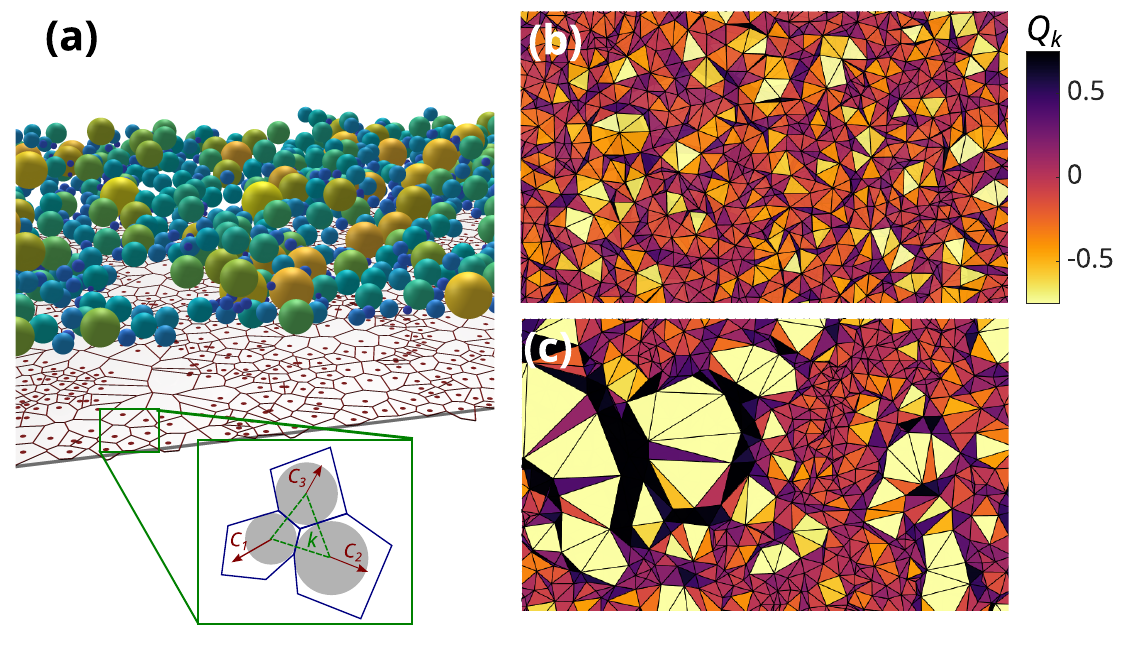}
  \end{center}
\caption{Characterizing the structural anisotropy of the deposited layer. (a) Granular packing of powder layer for $Bo=190.6$. The points represent the projections of particle centers on the $xy$-plane. The corresponding 2D Vorono\"{\i} tessellations are also shown on the plane. Also shown in (a), are a schematic representation of the particle packing with superimposed Vorono\"{\i} tessellation (blue) and Delaunay triangles (green). The vectors $C_p$ (red) point from particle centers to the centroids of Vorono\"{\i} cells. The calculated $Q_k$ for (b) $Bo=0$ and (c) $Bo=272.3$, where the Delaunay triangles are colored by the corresponding $Q_k$ values. }
\label{fig:voronoi}
 \end{figure}

\autoref{fig:voronoi}(a) shows an example of the deposited layer of highly cohesive particles $(Bo=190.6)$. We observe a heterogeneous structure comprising regions of dense and loose packing. At the particle scale, the density of neighbors surrounding each particle can be highly anisotropic, which could have important implications for subsequent processes such as heat transfer and phase change. Although such spatial fluctuations can be reflected by bin-averaged density, as done in \autoref{fig:packingFraction}, and the degree of fluctuation of densities at different locations depends on a manually chosen bin sizes. To avoid the need to manually specifying sampling length scales, we adapt a method of characterizing the structural heterogeneity using a particle-level measurement based on the packing anisotropy from the Vorono\"{\i} tessellation~\cite{rieser2016divergence}.
\subsection{Anisotropy Vector and Divergence}
The measure of the local structural anisotropy was developed by Rieser et al. from the observation that the center of a particle deviates from the centroid of its Vorono\"{\i} cell in a disordered packing \cite{richard2020predicting}. Any two particles with a shared Vorono\"{\i} cell face are defined as neighbors; from this, a Delaunay triangulation is generated by connecting groups of three mutual neighbors into triangles. \autoref{fig:voronoi}(a) includes a schematic illustration of the Vorono\"{\i} tessellation calculated based on the projections of the particle positions on the xy-plane, which is plotted below the particle packing with the Delaunay triangle $k$ and the anisotropy vector $\vec{C}$ pointing from particle centers to corresponding Vorono\"{\i} cell centroids. For a triangle representing a densely occupied area (overpacked, like the one depicted), all the $\vec{C}$ vectors point outward. For a triangle representing a void (underpacked), the vectors point inward. We quantify the extent to which the vectors  $\vec{C}$ at the three vertices of a Delauney triangle point inward or outward by calculating the divergence of the vectors of a triangle $k$ with area $A_k$. This is calculated based on the concept of constant strain triangle in finite element analysis \cite{cook2007concepts}. The local structural anisotropy, $Q_k$, calculated from the divergence, defined as

\begin{equation}\label{eq:Q_k^o}
 Q_k \equiv \left(\nabla \cdot \vec{C}\right)\,\frac{A_k}{\bar{A}}\,,
\end{equation}

\noindent where $\bar{A}$ is the average of all $A_k$ within the packing. By construction, $Q_k$ is dimensionless with a mean near zero. It is sensitive to the local structural anisotropy and has a geometrical significance: positive (negative) values correspond to overpacked (underpacked) regions. 

%The distribution of $Q_k^o$  values over a packing is nearly Gaussian \cite{rieser2016divergence}; hence, it is well described by the standard deviation and the skewness.

%\subsection{Modification for the Quasi-2D Deposited Layer}
The challenging aspect here is to extend the 2D calculation to a quasi-2D thin free-surface layer with a height of two to three particle diameters, which is typical in powder spreading. Since the deposited layer is thin and many regions only contain a single layer of particles, as shown in \autoref{fig:voronoi}, we use the 2D projections of the particles on the $xy$ plane for calculating the anisotropy. This simplification could lead to contributions of highly positive $Q_k$, especially in non-cohesive packing, due to vertically aligned particles in a quasi-2D layer. To see the influence of such scenarios,  we scale $Q_k$ with the ratio of the projected area of overlapping particles, $A_p$, on the $xy$ plane and the sum of the real area of the particles, $A_r = \sum\pi r_i^2$. The scaled $Q_k^\prime$ is given as follows:
\begin{equation}\label{eq:Q_k^o_mod}
Q_k^\prime \equiv  Q_k\,\frac{A_p}{A_r} = \left(\nabla \cdot \vec{C}\right)\,\frac{A_k}{\bar{A}}\,\frac{A_p}{A_r}\,.
\end{equation}
For highly dense packings of vertically aligned particles, $A_p/A_r < 1$, and the anisotropy is reduced. For dilute packings, $A_p/A_r = 1$, and thus $Q_k^\prime$ coincides with the original definition without correction.

\subsection{Divergence Fields and Distributions}
\autoref{fig:voronoi}(b) and (c) show the $Q_k$ map for $Bo=0$ and $Bo=272.3$, respectively, where the triangles are colored according to the corresponding values of $Q_k$. For $Bo=0$, the triangles share similar areas, and the $Q_k$ value fluctuates between positive and negative randomly in space. For $Bo=272.3$, large triangles corresponding to underpacked regions exist, making the nearby $Q_k$ values highly positive or negative, indicating strong anisotropy. Note that the $Q_k$ value of each individual triangle is determined by its immediate neighborhood, rather than the overall packing density. The dense and homogeneous regions in both $Bo=0$ and $Bo=272.3$ have low $Q_k$ values and only anisotropic regions show extreme values, which mostly exist in $Bo=272.3$. 
 %\begin{figure}[!htb]
 % \begin{center}
 %   \includegraphics[width=0.8\columnwidth]{figures/QkContour1.png}
 % \end{center}
%\caption{Delaunay triangles with color representing local values of $Q_k$ for (a) $Bo=0$, and (b) $Bo\approx30$ for a section of deposited layer $x \in (4\dots 6)$ mm and $y \in (0.2\dots 8)$ mm.}
%\label{Q_k}
% \end{figure}
 \begin{figure}[t]
  \begin{center}
  %\fbox
    {\includegraphics[width=\columnwidth]{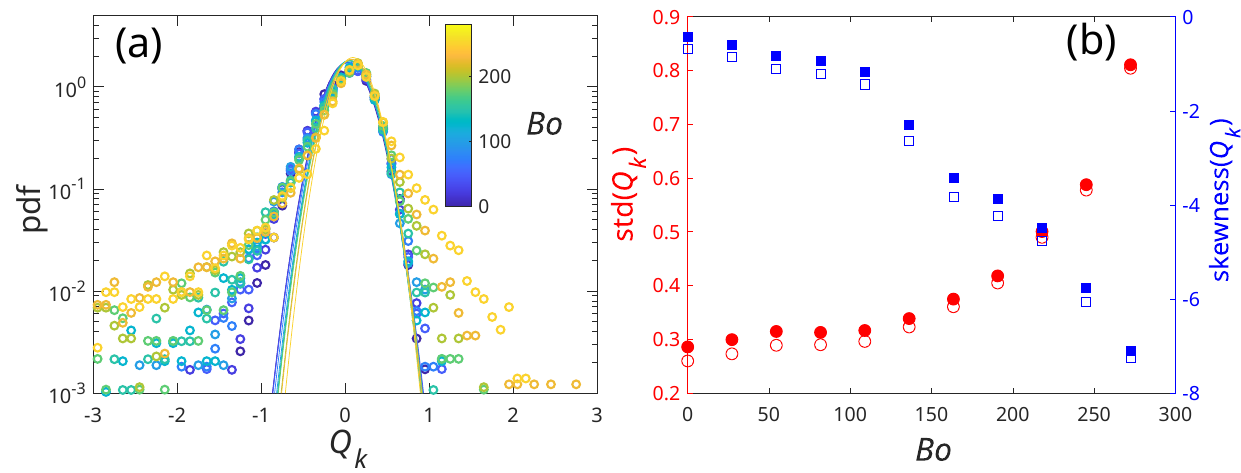}}
  \end{center}
  \caption{(a) Probability density of normalized divergence of center-to-centroid vectors for the quasi-2D packing of powder deposited for different $Bo$. $Q_k  > 0$ regions are more densely packed than their surroundings; hence, we call these regions overpacked. $Q_k < 0$ regions are more loosely packed than their surroundings and are, therefore, labeled underpacked. The solid curves are Gaussian fits $\bar{Q}_k - 0.5$ to $\bar{Q}_k + 0.5$. (b) Standard deviations (red circles) and skewness (blue squares) vs. $Bo$. The standard deviations and skewness of the distributions of  $Q_k$ (solid) and $Q_k^\prime$ (hollow) are compared.}
\label{fig:pdf}
 \end{figure}

The distribution of $Q_k$ is a strong structural indicator that is associated with important mechanical properties of a disordered packing, including jamming and shear band formation \cite{rieser2016divergence,xiao2020strain,harrington2018anisotropic,harrington2020stagnant}. In dense and homogeneous regions, $Q_k$ fluctuates randomly, and a peak in the $Q_k$ distribution around $Q_k=0$ is typically observed \cite{rieser2016divergence}. In heterogeneous regions with high anisotropy, the highly positive and negative $Q_k$ values show up together on the tails of the distribution, making them deviate from Gaussian. \autoref{fig:pdf}(a) shows the distribution of $Q_k$ for different $Bo$. The majority of the $Q_k$ resides in the region around zero. It can be fitted to a Gaussian distribution using values between $\bar{Q}_k - 0.5$ to $\bar{Q}_k + 0.5$ for each data set (solid curves), where $\bar{Q}_k$ is the mean of the distribution. For lower $Bo$, we observe a consistent slope of the distribution throughout the range $Q_k < 0$, which is an indication of a homogeneous structure throughout the packing. In contrast, a transition of the slope of the distribution at $Q_k = -1$ is clearly distinct for higher $Bo$, suggesting the coexistence of dense homogeneous regions and dilute heterogeneous regions for highly cohesive materials: for $Q_k < -1$, the distribution deviates from Gaussian and becomes exponential-like for higher $Bo$. This exponential tail corresponds to the existence of highly underpacked sites distributed sparsely in the packing, as seen in \autoref{fig:voronoi}(c); for $-1 < Q_k < 1$, the distribution is narrower with increasing $Bo$, indicating the existence of locally homogeneous packing. This variation in structure is also evident in cohesive systems shown in the experimental studies by Xiao et al. \cite{xiao2020strain}. Such a variation is not observed for non-cohesive experimental particle systems in Harrington et al. \cite{harrington2020stagnant} for disordered particle packings. 
 
To quantify the difference in packing heterogeneity for different $Bo$, we show the standard deviation and the skewness of the $Q_k$ distribution in \autoref{fig:pdf}(b). The standard deviation reflects the portion of highly anisotropic sites (triangles) in a packing. The skewness roughly compares the degree of anisotropy of loosely packed sites to densely packed sites. 
%and observe a transition behavior of these parameters at $Bo=10$. 
%The standard deviation (skewness) remains constant up to $Bo\approx10$ and a monotonic increase (decrease) for $Bo>10$. Thus, $Bo\approx10$  shows a transition in structural anisotropy for cohesive materials. 
For higher cohesion, particles can sustain more voids during spreading and encounter higher local anisotropy, leading to a more heterogeneous overall packing structure. As a result, the standard deviation increases with $Bo$, which reflects the difference seen in \autoref{fig:voronoi} in a quantitative way. The skewness decreases with $Bo$, which reflects the growing tail at the negative end of the $Q_k$ distribution. This corresponds to the fact that the void sites not only grow larger in number, but also have larger sizes at higher $Bo$.
%We observe fluctuations between repetitions of the simulations in the trend for higher $Bo$ which is a result of the high anisotropy of the structures. 
We compared the standard deviations and the skewness of the distributions with the anisotropy calculated using \autoref{eq:Q_k^o} and \autoref{eq:Q_k^o_mod} for $Q_k^\prime$, respectively. The divergence $Q_k$ from the condition in \autoref{eq:Q_k^o} gives slightly higher values for both the standard deviation and the skewness but qualitatively shows the same behavior as $Q_k^\prime$.
\section{Surface roughness characterization of powder layer}

\subsection{Digitized free surface height characterization}

\begin{figure}[t!]
%\begin{adjustwidth}{-\extralength}{0cm}
\centering
%\fbox
{\includegraphics[width=\textwidth,trim={5.0cm 1.6cm 2.5cm 2cm},clip]{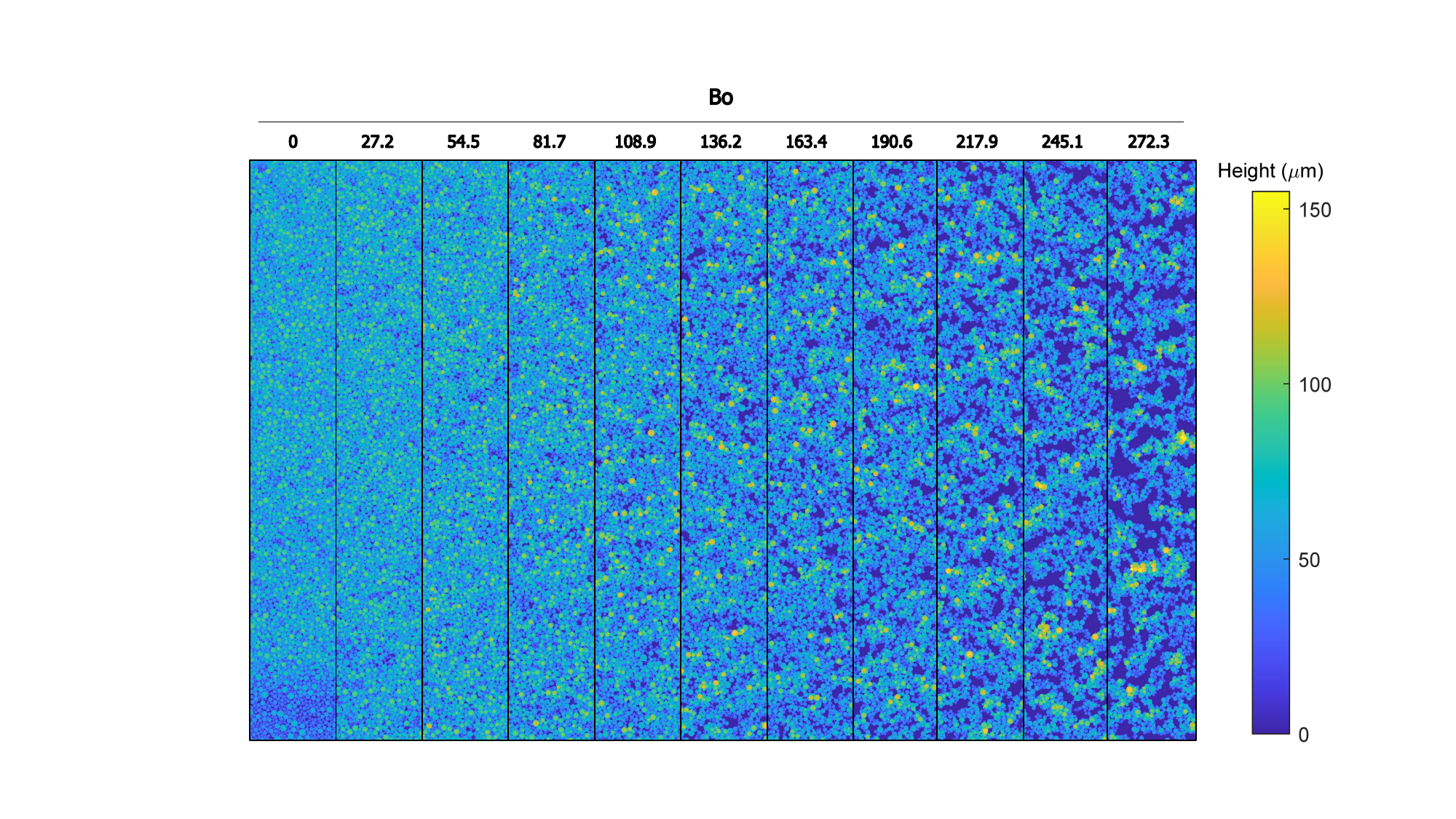}}
%\end{adjustwidth}
\caption{Height of powder layer surfaces for various Bond numbers from $Bo=0$ to $Bo=272.3$.
%, measured from the level of the spreading surface.
\label{fig:powder_layers}}
\end{figure}

%Powder spreading is a serialized, multi-step additive manufacturing process, where each consecutive layer is deposited on top of the previous one. 
Measuring the surface roughness of the powder layer is of interest, as it is related to the roughness of the final product, while distinct rough features are areas prone to become the source of defects. As discussed in the previous sections, packing density and structural packing anisotropy are integral elements in assessing the quality of the finished part. However, they do not provide information on the irregularity of the surface texture of the powder layer. To this end, it is useful to characterize the surface roughness of the produced layers corresponding to different cohesion values. Various aspects of roughness can be characterized via the calculation of independent quantitative indices that provide diverse morphological information on the layer's topography. 

\begin{figure}[t]
  \begin{center}
  %\fbox
    {\includegraphics[width=1.0\columnwidth]{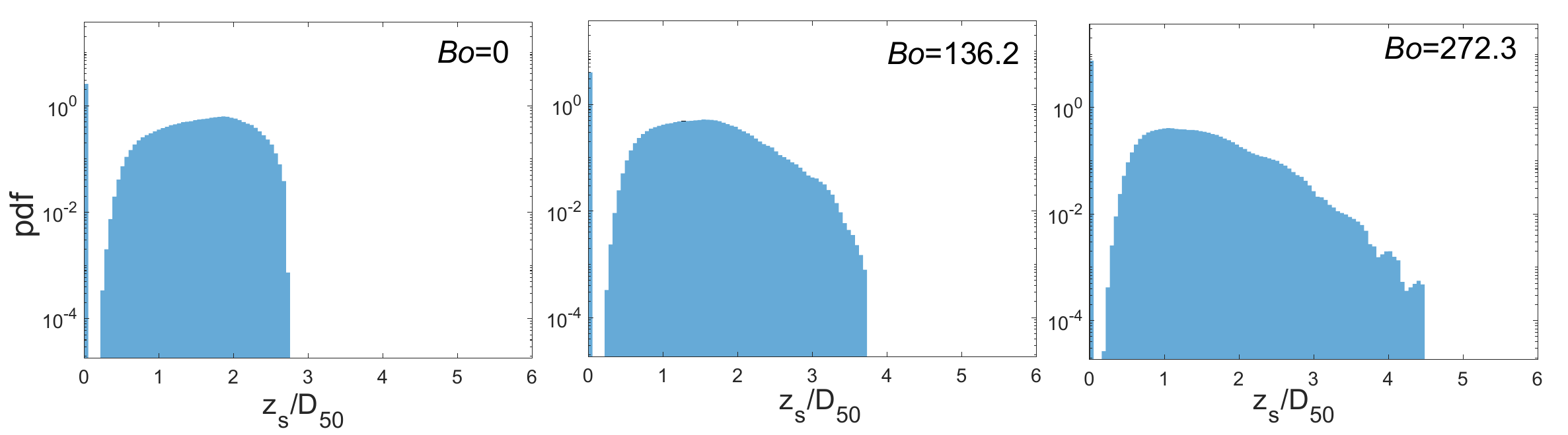}}
  \end{center}
\caption{The distribution of digitized free surface height distributions for $Bo=0$, $Bo=136.2$, and $Bo=272.3$. }
\label{fig:hdist}
 \end{figure}

Previous work focused on the two-dimensional characterization of surface roughness features of powder layers, using indices that correspond to planar rough profiles, usually taken as representative of the real rough profile \cite{parteli2016particle,nasato2021influence}.
%\TP{What does the preceding sentence say?} 
Here, the three-dimensional surface profile of each powder layer was reconstructed for each layer after spreading is completed. The particles located near the top of the powder layer were identified, and points on their surface were calculated using a regular sampling grid~\cite{angelidakis2021shape}, giving a surface height profile, $z_\text{s}$, similar to Meier et al.~\cite{meier2019critical}. Note that if the surface of the substrate is directly exposed at a sampling point, a value of zero is recorded.
\autoref{fig:powder_layers} shows the surface heights of the 11 studied powder layers of different cohesion. Interestingly, the surface height reaches a maximum value of up to $150\,\mu$m for larger Bond numbers, which is larger than the gap height of $100 \,\mu$m, shown in \autoref{fig:schematic}(b). This typically occurs for fine cohesive powders, due to decreased flowability of the powders when the cohesion effects dominate gravity and inertia, resulting in the formation of agglomerates with internal cavities and irregular surface profiles \cite{chen2019powder,ma2020numerical}. The surface height of the cohesionless powder layer ($Bo=0$) does not exceed the gap height. \autoref{fig:hdist} shows three example distributions of the measured surface height, which shows that the distribution widens as $Bo$ increases. However, the shapes of the height distributions are rather complicated, and require many parameters to describe as listed in the following subsection. A spike at $z_s=0$ exists for all three $Bo$, which corresponds to the exposed substrate surface.

\subsection{Roughness characterization using height distributions}

\begin{figure}[htb!]
    \label{fig:height}

    \centering
     \begin{minipage}{0.5\textwidth}
        \centering
        %\fbox
        {\includegraphics[width=1.0\textwidth,trim={3cm 9cm 3cm 9cm},clip]{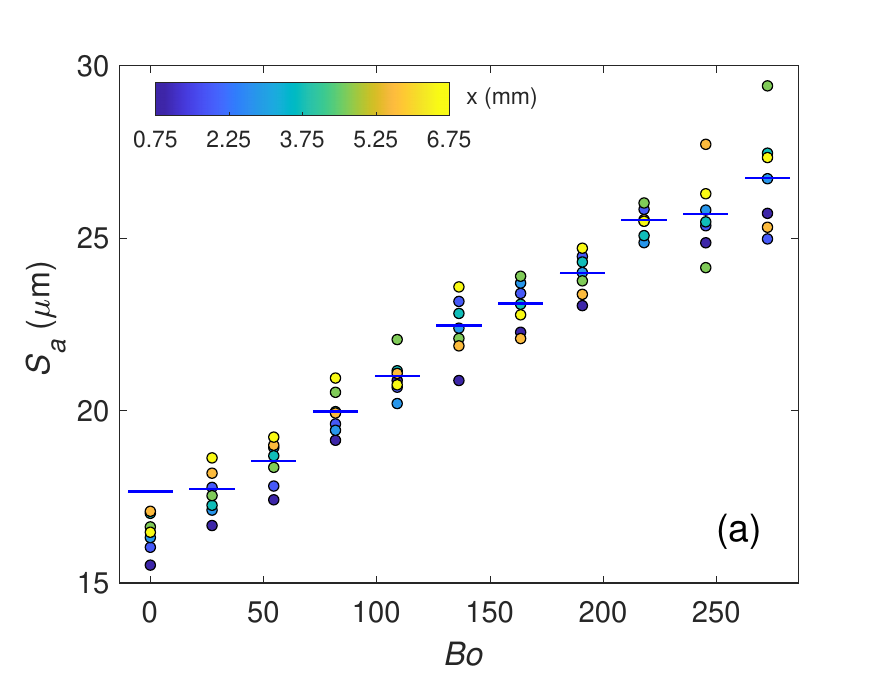}}
        \label{fig:roughness:sa}
 \end{minipage}\hfill
     \begin{minipage}{0.5\textwidth}
        \centering
        %\fbox
        {\includegraphics[width=1.0\textwidth,trim={3cm 9cm 3cm 9cm},clip]{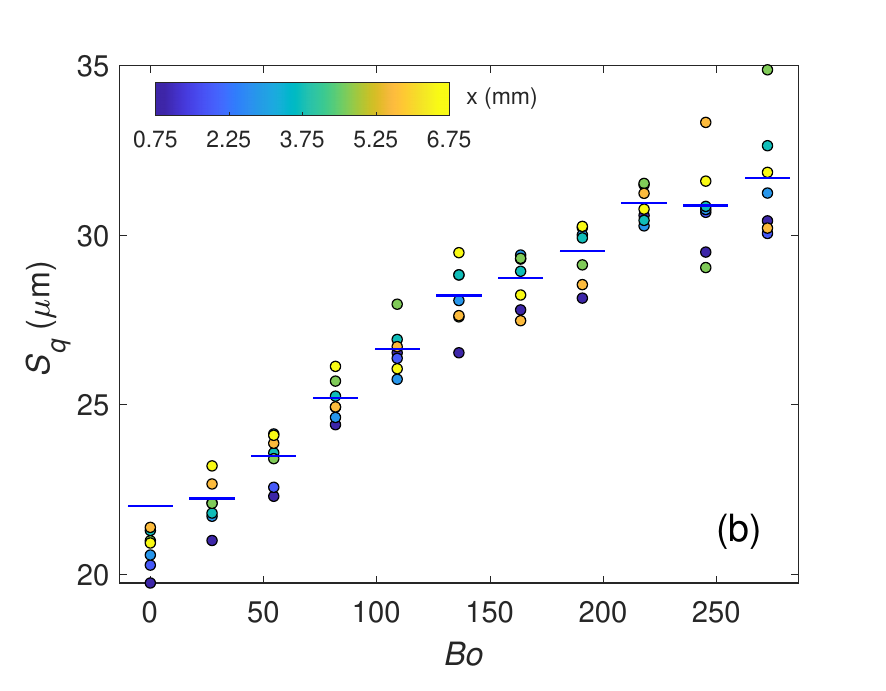}}
        \label{fig:roughness:sq}
 \end{minipage}\hfill
      \begin{minipage}{0.5\textwidth}
        \centering
        %\fbox
        {\includegraphics[width=1.0\textwidth,trim={3cm 9cm 3cm 9cm},clip]{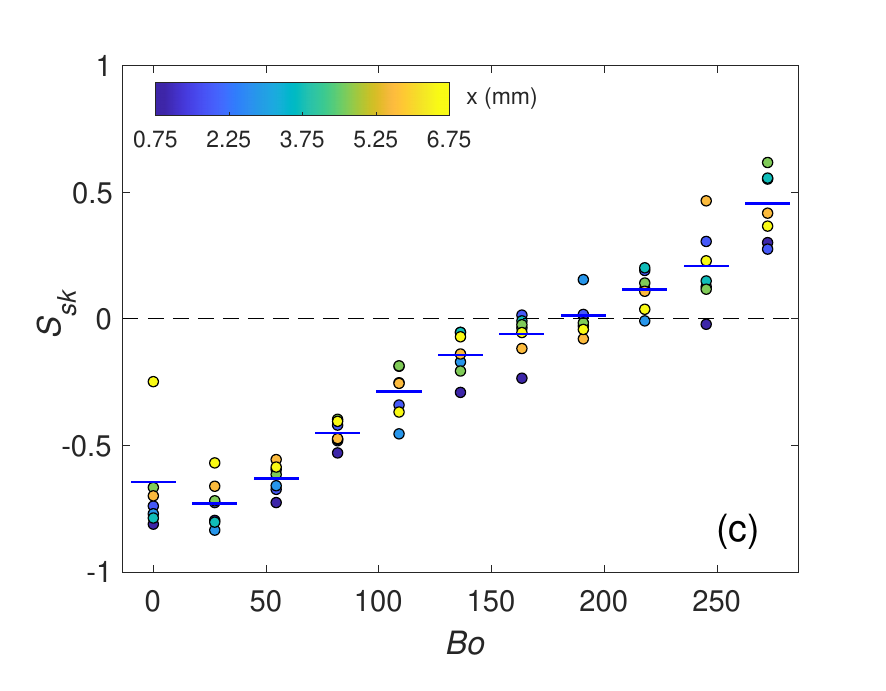}}
        \label{fig:roughness:ssk}
 \end{minipage}\hfill
       \begin{minipage}{0.5\textwidth}
        \centering
        %\fbox
        {\includegraphics[width=1.0\textwidth,trim={3cm 9cm 3cm 9cm},clip]{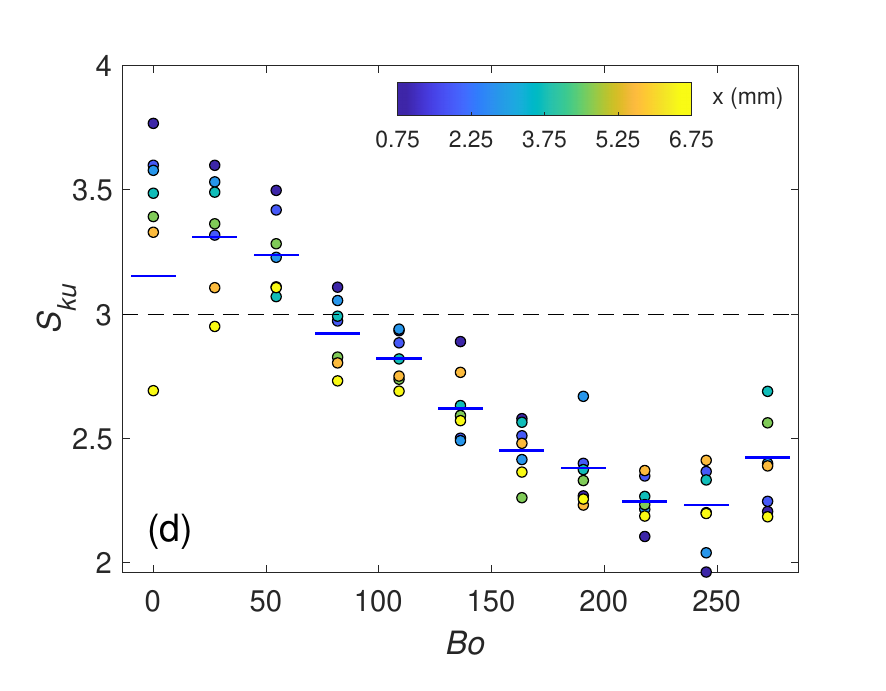}}
        \label{fig:roughness:sku}
 \end{minipage}\hfill
\caption{Surface roughness parameters (a) arithmetic mean height $S_a$ (b) root mean square height $S_q$ (c) skewness $S_{sk}$ and (e) kurtosis $S_{ku}$ shown as a function of $Bo$. The solid horizontal lines note the global values of the surface roughness parameters for each layer. The sample points are colored according to their distance $x$ from the starting point of spreading. The dashed line for $S_{sk}=0$ marks the threshold between profiles where most rough features appear above the mean plane ($S_{sk}<0$) and below it ($S_{sk}>0$). The dashed line for $S_{ku}=3$ marks the threshold between rough profiles with rounded peaks ($S_{ku}<3$) and with sharp ones ($S_{ku}>3$). %{\color{red}{@Vasilis: Please mention a sentence about the indication of the dashed lines? Sudeshna} \textbf{Good find, done.}}
}\label{fig:surfaceRoughness}
\end{figure}

The current state-of-the-art for characterizing rough surfaces, as outlined in ISO 25178 \cite{ISO25178}, calculates roughness indices based on the surfaces of real, three-dimensional texture profiles. Using distributions of $z_{\text{s}}$, the surface roughness is characterized in terms of arithmetic mean height $\left(S_a\right)$, root mean square height $\left(S_q\right)$, skewness $\left(S_{sk}\right)$ and kurtosis $\left(S_{ku}\right)$.
%The indices employed to characterize surface roughness (i.e. $S_a$, $S_q$, $S_{sk}$, $S_{ku}$) represent 
The height deviation of the surface roughness from a mean surface height of the entire layer is used, which is $z_m$. We show these roughness parameters in \autoref{fig:surfaceRoughness} for increasing Bond number values, where the points are colored according to their distance $x$ from the start of spreading, and along the spreading direction. A clear correlation was not found between any of the surface roughness parameters and their distance from the initial spreading position. We next discuss the significance of the roughness parameters individually.
The arithmetic mean height is calculated as:

\begin{equation}\label{eq:sa}
    \displaystyle S_a = \frac{1}{A} \iint\limits_A \left|\bar{z} \left(x,y \right) \right| \diff x \diff y.
\end{equation}

where $\bar{z}=z_{\text{s}}-z_m$ is the height of a point on the layer surface, measured from the plane of mean surface height, $z_{\text{s}}$ is the measured free surface height, $x$ and $y$ the horizontal coordinates of the point along and transversely the spreading direction, and $A$ the area occupied by the layer. It becomes evident in \autoref{fig:surfaceRoughness}(a) that for powders of increasing cohesion, the arithmetic mean height increases almost linearly with the Bond number up to values of $Bo = 217.9$. This indicates that cohesive powders lead to higher deviations from the mean height, and to rougher surface texture profiles. Also, the scatter of measurements increases slightly for larger Bond numbers, which points to the conclusion that more cohesive powders feature more heterogeneous profiles, with taller peaks and deeper valleys.
To further validate this trend, the root mean square height is calculated as:

\begin{equation}\label{eq:sq}
    \displaystyle S_q = \sqrt{\frac{1}{A} \iint\limits_A \bar{z}^{2}\left(x,y\right)  \diff x \diff y}.
\end{equation}

It can be seen in \autoref{fig:surfaceRoughness}(b) that the root mean square height presents the same general trend as the scattering of the arithmetic mean, where more cohesive powders form layers with more heterogeneous height distributions.
%This parameter demonstrates a higher degree of scatter for more cohesive powders than the arithmetic mean.
This is in agreement with findings from the literature \cite{meier2019critical,he2020linking}. These two measures of the average height of the powder surface texture are informative regarding the extent of the roughness, but provide no information about their morphology. To this end, the skewness and kurtosis of the surface height profiles are examined.
The height skewness is calculated as:

\begin{equation}\label{eq:ssk}
    \displaystyle S_{sk} = \displaystyle \frac{1}{{S_q}^{3}} \left[ \frac{1}{A} \iint\limits_A \bar{z}^{3} \left( x,y \right) \diff x \diff y \right].
\end{equation}

Skewness is a measure of the asymmetry of the layer height distribution around the mean plane. Negative skewness values ($S_{sk}<0$) indicate that the height distribution is skewed above the mean height plane, with a few deep valleys, zero skewness values ($S_{sk}=0$) correspond to a symmetric surface, where peaks and valleys occupy the same amount of surface in average, while positive values ($S_{sk}>0$) indicate that the height distribution is skewed below the mean height plane, with a few tall peaks. \autoref{fig:surfaceRoughness}(c) shows a monotonically increasing trend of skewness with increasing Bond number values, where powders with lower cohesion ($Bo<190.6$) demonstrate negative skewness, with average cohesion ($Bo\approx190.6$) nearly zero skewness and with higher cohesion ($Bo>190.6$) positive skewness values. 
%The terms smaller, average and higher cohesion here, reflect the range of the studied Bond numbers and are not meant to be a generalized comment on how cohesive each examined powder material is in a more general way. %{\color{red}++ Interpretation/Hypothesis/Mechanism?}
The height kurtosis is calculated as:

\begin{equation}\label{eq:sku}
    \displaystyle S_{ku} = \displaystyle \frac{1}{{S_q}^{4}} \left[ \frac{1}{A} \iint\limits_A \bar{z}^{4} \left( x,y \right) \diff x \diff y \right].
\end{equation}

Like skewness, kurtosis describes a particular morphological aspect of the surface height distribution. Skewness is a metric of whether most of the rough profile is positioned above or below the mean height plane. Kurtosis provides information on the average shape of the surface texture asperities, and can be seen as a probability density sharpness of the rough features. Low kurtosis values ($S_{ku}<3$) indicate platykurtic surface texture profiles of well-rounded asperities presenting short tails, zero kurtosis values ($S_{ku}\approx3$) correspond to mesokurtic profiles of Gaussian-like asperities characterized by medium-sized tails, while high kurtosis values ($S_{ku}>3$) correspond to leptokurtic surface profiles, with sharp, spike-like asperity characteristics presenting long tails.

\autoref{fig:surfaceRoughness}(d) shows the kurtosis values for the various powder layers of varying cohesion, where the parameter shows a non-monotonous, mostly declining trend for increasing Bond number. It becomes evident that the powder layer corresponding to zero cohesion ($Bo=0$) features high kurtosis values ($S_{ku}>3$), while layers made of cohesive powders feature lower kurtosis values ($S_{ku}<3$). For the higher end of the studied cohesion levels ($Bo>217.9$) kurtosis shows a mild increasing trend, which is however characterized by a high degree of scatter, making a further interpretation challenging. 
%{\color{red}++ Interpretation/Hypothesis/Mechanism?}

Combining the observations of all roughness parameters for the studied powder layers of varying Bond number, it can be inferred that increasing cohesion leads to powder layers characterized by increased roughness, where the layer lies mostly below its average height, and presents a few, rounded peaks. For less cohesive powders, the corresponding layers are characterized by less pronounced rough features of a sharper nature. These observations can possibly be explained by considering that cohesive particles tend to agglomerate into larger clusters, which appear to be more rounded at the scale of the full powder layer, compared to cohesionless particles which pack without demonstrating clustering, and thus it is more probable for them to have individual particles deposited on the surface of the layer, which macroscopically resemble sharp peaks.

\subsection{Spatial fluctuation of the free surface height}

\begin{figure}[t]
  \begin{center}
  %\fbox
    {\includegraphics[width=0.85\columnwidth]{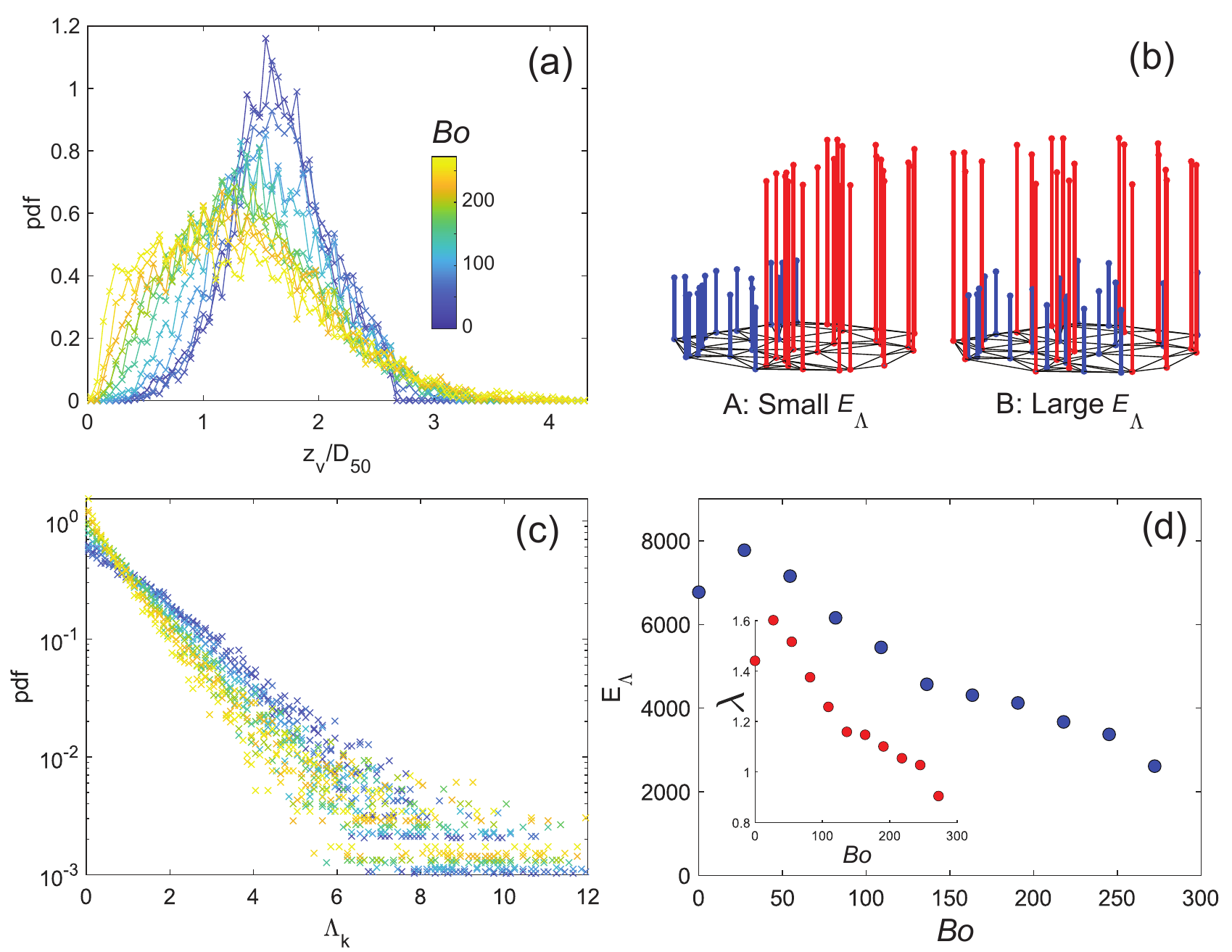}}
  \end{center}
\caption{Quantifying mixing of surface heights. (a) Distributions of Vorono\"{\i} cell-averaged surface height for different $Bo$. (b) Illustration of poor mixing (left) and well mixing (right) that generate the same surface height distribution. (c) Distributions of the Dirichlet energy of individual triangles for different $Bo$. (d) The total Dirichlet energy for each $Bo$. Inset shows the fitted exponential distribution constant for each $Bo$.  }
\label{fig:mixing}
 \end{figure}

While examining the digitized height distribution is informative, it does not contain information on the spatial arrangement of the height profile. For a given set of digitized height values, a permutation of their spatial arrangement does not change the distribution. This problem is analogous to the problem where a global packing density does not offer information on the homogeneity of the packing. Therefore, we again use the projection-based Vorono\"{\i} and Delaunay tessellations as in Section \ref{sec:local_structural_anisotropy} to address this issue. For a sphere packing, the digitized free surface height values for each sphere are spatially correlated as they can be fully described by the center coordinates and the radius of the sphere. To reduce this correlation, a spatial coarse graining at the length scale of a particle's diameter is required. We average the free surface height value, $z_{\text{s}}$, in each Vorono\"{\i} cell, defining a cell-averaged height, $z_v$, at the center of each corresponding sphere, which is also a vertex in the Delaunay triangulation. The calculated distributions of $z_v$ for different cohesion are shown in
\autoref{fig:mixing}(a), which are colored by the corresponding $Bo$. Results show that the distribution widens with increasing $Bo$, which agrees with results in \autoref{fig:surfaceRoughness}. The variance of the $z_v$ distribution can be calculated as $\sigma_{z_v}^2$, but as mentioned earlier, it does not contain information of the spatial height fluctuation.

To demonstrate the permutation problem, a sketch is made in \autoref{fig:mixing}(b) where the Delaunay triangles are drawn in black, and the height of each vertical stick from a vertex represents $z_v$. For simplicity, we show idealized scenarios with only two height values, which can be organized into scenario A where the short surfaces (blue) and tall surfaces (red) are spatially segregated, and scenario B where they are mixed, with both cases having the same height distribution. The degree of ``mixing" between taller and shorter surfaces needs to be quantified for a more complete description of the free surface profile. This can be described by how different the values are for the three vertices in an triangle, and this difference is small for most triangles in A and large for most triangles in B. To quantify this, we use the square of the first spatial derivative, $|\nabla z_v|^2$, and an integration of this quantity over the entire domain, which is essentially the Dirichlet Energy \cite{pinkall1993computing,chen2010spectral,ye2020dirichlet}
\begin{equation}\label{eq:DE}
    E_\Lambda = \frac{1}{2\sigma^2_{zv}} \int |\nabla z_v|^2dA,
\end{equation}

\noindent which quantifies the degree of variation of a function in a given domain, with the function being the height, $z_v$, that varies on the 2D domain $A$ on the $xy$ plane. In a lattice triangulation, this quantity can be digitized as

\begin{equation}\label{eq:DE_d}
    E_\Lambda = \frac{1}{2} \sum_{k} \Lambda_k = \frac{1}{2} \sum_{k(l,m,n)} \frac{1}{2\sigma^2_{zv}}\left[ \cot{\alpha_{lm}}(z_{v,l}-z_{v,m})^2 + \cot{\alpha_{ln}}(z_{v,l}-z_{v,n})^2 + \cot{\alpha_{mn}}(z_{v,m}-z_{v,n})^2\right],
\end{equation}

\noindent where $\Lambda_k$ is the normalized Dirichlet energy for a single triangle $k$, and $l,m,n$ are the vertices of $k$, and $\alpha_{lm}$ is the angle facing the edge connected by $l$ and $m$. The distribution of $\Lambda_k$ of all analyzed triangles for each $Bo$ is shown in \autoref{fig:mixing}(c). For each $Bo$, the distribution is a straight line on a log-lin scale suggesting an exponential distribution, $P(\Lambda_k)=\lambda e^{-\lambda\Lambda_k}$. Unlike the distribution of the digitized surface height with complicated shapes and spikes at $z_\text{s}=0$ (\autoref{fig:hdist}), the exponential distribution can be conveniently described by a single parameter, $\lambda$, which sets the rate of decay for $P(\Lambda_k)$. It can be seen from \autoref{fig:mixing}(c) that the more cohesive cases have faster decays with higher values near zero.

To quantify the variation and the decay, we plot the total Dirichlet Energy for each $Bo$ in \autoref{fig:mixing}(d) and the fitted distribution parameter $\lambda$ as an inset, both decreasing with $Bo$. Note that with the normalization by $\sigma_{z_v}^2$, these two quantities truly reflect the blending of taller and shorter surfaces, not the spread of surface heights. These results quantitatively show that at the length scale set by particle size, higher cohesion results in less local height fluctuation, despite having a higher spread in height values. This is because low cohesion particles pack densely and homogeneously, and the surface height fluctuates at the particle scale, which is similar to scenario A in \autoref{fig:mixing}(a). On the other hand, high cohesion particles form dilute and heterogeneous packings with clustering that is more similar to scenario B. In this sense, the local surface height fluctuations and the local packing anisotropy in a layer should be closely related, which is subject to future studies.

\section{Conclusions}
This work quantifies the structural features with a focus on density and surface roughness in powder layers in DEM simulations using realistic cohesive interaction forces. We first used a more traditional approach by calculating global values to show the general trend of decreasing density and increasing surface roughness as cohesion increases. The global structural features was calculated by digitization of the simulated spheres at a very fine scale and then samples globally by binning for density and by examining the distribution for the surface height profile. The increase in the surface roughness was then further interpreted by examining higher moments of the height distribution, including the skewness and the kurtosis, both show a gradual evolution for layers with neighboring Bond numbers. In particular, for $Bo=0$ the skewness $S_{sk}<0$ and kurtosis $S_{ku}>3$, indicating that most rough features appear above the mean height plane and have sharp peaks, while for $Bo=272.3$ we observe the inverse trend, i.e. the skewness $S_{sk}>0$ and kurtosis $S_{ku}<3$, indicating that most rough features appear below the mean height plane and have more rounded peaks. %The region around $Bo\approx10$ marks the transition of these features of surface roughness, which is also the bifurcation point for the structural anisotropy index. 

%The packing density showed a gradual evolution for increasing Bond number, without a clear transition point. Cohesive powders were found to lead to layers of lower packing fraction and increased local density inhomogeneity. Regarding the latter, different sections of the same layer demonstrated a moderate range of densities, a trend that became more profound for powders of increasing Bond number.

To highlight the increasing heterogeneity of the density and surface profile, we also developed Vorono\"{\i}-based metrics that quantifies the spatial fluctuations of these quantities of interest.
For density fluctuation, the divergence of the Vorono\"{\i} anisotropy vector, $Q_k$, was adopted for the thin deposited particle layers as a geometrical measure of their structural heterogeneity. The transition in the slope of $Q_k$ distribution at $Q_k = -1$ displays a signature for the coexistence of dense regions with homogeneous structures as well as dilute regions with highly anisotropic structures, which is typical for cohesive materials. With increasing cohesion, both the standard deviation and skewness of the $Q_k$ distributions exhibit a consistent, monotonic change, indicating increasing structural heterogeneity of the deposited layer.

%The surface roughness parameters showed milder variations for increasing Bond number, without a clear transition, where cohesive powders were found to form layers of increased surface roughness. Interesting to note that skewness and kurtosis showed a gradual evolution for layers with neighboring Bond numbers, but a dramatic overall increase if we compare the layers made of cohesionless ($Bo=0$) and very cohesive ($Bo=272.3$) materials. In particular, for $Bo=0$ the skewness $S_{sk}<0$ and kurtosis $S_{ku}>3$, indicating that most rough features appear above the mean height plane and have sharp peaks, while for $Bo=272.3$ we observe the inverse trend, i.e. the skewness $S_{sk}>0$ and kurtosis $S_{ku}<3$, indicating that most rough features appear below the mean height plane and have more rounded peaks. %The region around $Bo\approx10$ marks the transition of these features of surface roughness, which is also the bifurcation point for the structural anisotropy index. 
%Thus, we argue that there is a gradual transformation of the structural anisotropy and the surface roughness profile with increasing cohesion, from $Bo=0$ to $Bo=272.3$.

We quantified the fluctuation of the free surface height using the Vorono\"{\i} cell-averaged height. Instead of focusing on the global distribution of this height, which contains no information on the spatial arrangement of the height values, we calculate the local squared spatial gradient as a measure of how well the taller and shorter surfaces are mixed. The distribution of the squared gradient is exponential which can be quantified by a single parameter. When normalized by the variance of the surface height, both the fitted distribution parameter and the total sum of the squared gradient show a decrease with increasing $Bo$. This quantitatively demonstrates that higher cohesion leads to reduced local height fluctuation despite the height values having a wider spread, which is possibly because that the packing density heterogeneity results in significant fluctuations of the free surface at a larger length scale. In contrast, at lower cohesion levels, particles densely and homogeneously pack, resulting in more surface height fluctuation at the particle scale.

The additional sets of metrics for the spatial fluctuation of density and surface height, combined with the global metrics, offer a more complete 
%\TP{(cannot be complete for simple reasons: The distribution is described by many numbers (the frequencies), but the moments are only a few numbers. There must be missing information, unless you make further assumptions.)}
description of the packing structure than the traditionally used bulk-averaged values. This set of parameters can not only serve as a quantification of the quality of spreading but can also be used as a structural basis for modeling and analysis of subsequent processes, such as heat transfer and binder infiltration, as the heterogeneity of the packing structure on the particle level is important in these processes. For thicker layers and non-spherical particles, the 2D projection-based Vorono\"{\i} calculation could lose its validity, but the same concept can be extended using real 3D set Vorono\"{\i} analysis~\cite{schaller2013set,phua2023understanding}.

%In conclusion, this study on powder spreading process and powder cohesion has wide-ranging implications for real-world applications. The arrangement of powder particles within each printed layer and the surface roughness can significantly impact the mechanical properties and structural integrity of the final printed part. For instance, controlled surface roughness in automotive components can reduce friction and wear, leading to more durable and efficient engines. Additionally, optimizing surface height fluctuations in critical areas like piston rings and bearings can extend the lifespan of these parts. Moreover, surface height fluctuation control in external automotive components such as spoilers and aerodynamic panels can improve aerodynamic efficiency, reducing drag and enhancing fuel efficiency for greener and more economical vehicles. The insights gained from this research also contribute to advancing various industries by optimizing powder manufacturing processes and tailoring material properties and surface characteristics. This, in turn, paves the way for enhanced product design and performance.

%\par
%This study demonstrates the value of combining different characterization methods to better understand additive manufacturing processes involving granular materials. This study is a step further towards achieving simulation-based optimization of powder spreading processes, via advanced analysis of the produced layers, for powders with a wide variety of properties. 

\section*{Acknowledgement}
We gratefully acknowledge Deutsche Forschungsgemeinschaft (DFG, German Research Foundation) for funding the Collaborative Research Center 814 (CRC 814), Project Number 61375930-SFB 814 `Additive Manufacturing', sub-project B1. We also thank Humboldt Research Foundation for granting the `Humboldt Research Fellowship'. The work was supported by the Interdisciplinary Center for Nanostructured Films (IZNF), the Competence Unit for Scientific Computing (CSC), and the Interdisciplinary Center for Functional Particle Systems (FPS) at Friedrich-Alexander-Universit\"at Erlangen-N\"urnberg.






%\bibitem{

%}
\bibliography{structural-cohesive}

\end{document}